\documentclass[english,a4paper]{article}
\usepackage[T1]{fontenc}
\usepackage[cp1250]{inputenc}
\usepackage{amssymb}
\usepackage{hyperref}
\usepackage{amsfonts}
\usepackage{graphicx}

\def\openone{\leavevmode\hbox{\small1\kern-3.3pt\normalsize1}}
\begin{document}

\title{External constraints on optimal control strategies in molecular orientation and photofragmentation: Role of zero-area fields}

\author{D. Sugny\footnote{Laboratoire Interdisciplinaire
Carnot de Bourgogne (ICB), UMR 5209 CNRS-Universit\'e de
Bourgogne, 9 Av. A. Savary, BP 47 870, F-21078 DIJON Cedex,
FRANCE, dominique.sugny@u-bourgogne.fr}, S. Vranckx\footnote{Service de Chimie Quantique et Photophysique, CP 160/09 Universit\'e Libre de Bruxelles, B-1050 Brussels, Belgium}, M. Ndong, O. Atabek\footnote{Institut des Sciences Mol\'eculaires d'Orsay (ISMO), B\^at. 350 UMR 8214 CNRS-Universit\'e Paris-Sud, Orsay, France}, M. Desouter-Lecomte\footnote{Laboratoire de Chimie Physique (LCP), UMR 8000 CNRS-Universit\'e Paris-Sud, Orsay, France}}

\maketitle

\begin{abstract}
We propose a new formulation of optimal and local control algorithms which enforces the constraint of time-integrated zero-area on the control field. The fulfillment of this requirement, crucial in many physical applications, is mathematically implemented by the introduction of a Lagrange multiplier aiming at penalizing the pulse area. This method allows to design a control field with an area as small as possible, while bringing the dynamical system close to the target state. We test the efficiency of this approach on two control purposes in molecular dynamics, namely, orientation and photodissociation.
\end{abstract}

\section{Introduction}
In recent years, advances in quantum control have emerged through the introduction of appropriate
and powerful tools coming from mathematical control theory \cite{pont,bryson} and by the use of sophisticated
experimental techniques to shape the corresponding control fields \cite{rice,brumer,revuerabitz}.
All these efforts have lead nowadays to an unexpected and satisfactory agreement between theory
and experiment. On the theoretical side, one major tool to design the control field is Optimal Control
Theory (OCT) \cite{pont,bryson}. Over the past few years, numerical iterative methods have been developed
in quantum control to solve the optimization problems. Basically, they can be divided into two families,
the gradient ascent algorithms \cite{glaser} and the Krotov \cite{krotov,reich} or the monotonic \cite{zhu,maday} ones.
The design of optimal control fields by standard iterative algorithms \cite{revuerabitz} can require the computation of
several hundreds numerical propagations of the dynamical quantum system. While the efficiency of this procedure has been
established for low dimensional quantum systems, this approach can  numerically be prohibitive for large dimensions.
In this latter case, it is possible to use more easily accessible numerical methods, such as the local control
approach \cite{revuelocal,revuealessandro}. Roughly speaking, the optimization procedure is built from a Lyapunov function
$\mathcal{J}^\mathrm{lc}$ over the state space, which is minimum (or maximum) for the target state. A control field that ensures
the monotonic decrease (or increase) of $\mathcal{J}^\mathrm{lc}$ is constructed with the help of the first derivative
$\dot{\mathcal{J}}^\mathrm{lc}$ of $\mathcal{J}^\mathrm{lc}$. Note that this approach has largely been explored in quantum control \cite{sugawara,wang}.
Thanks to the progresses in numerical optimization techniques, it is now possible to design high quality control along with
some experimental imperfections and constraints. Recent studies have shown how to extend the standard optimization procedures
in order to take into account some experimental requirements such as the spectral constraints \cite{spectrum1,spectrum2}, the non-linear
interaction between the system and the laser field \cite{nonlinear1,nonlinear2}, the robustness against experimental errors
\cite{kobzar,zhang}. In view of experimental applications in quantum control, it is also desirable to design pulse sequences
with a zero global time-integrated area. Several works have pointed out the inherent experimental difficulties associated with the use of non-zero
area fields \cite{you,rabitz,sugny,ortigoso,lapert}, in particular for laser fields in the THz regime. Since the dc component of the
field is not a solution of Maxwell's equation, such pulses are distorted when they propagate in free space as well as through focusing optics. The standard optimization procedures do not take into account this basic requirement, designing thus non-physical control fields. In this framework, a question which naturally arises is whether one can adapt the known optimization algorithms to this additional constraint. This paper aims at taking a step toward the answer of this open question by proposing new formulations of optimal and local control algorithms. The zero-area requirement for the laser fields is mathematically fulfilled by the introduction of a Lagrange multiplier and the derivation of the corresponding optimal equations.

The goal of this paper is to explore the efficiency of the generalized optimal and local control algorithms on two key control problems
of molecular dynamics: orientation and photodissociation.
The remainder of this paper is organized as follows. The new formulations of optimization control algorithms are presented
in Sec. \ref{sec2}. Section \ref{sec3} is devoted to the application of the optimal control algorithm to the enhancement of
molecular orientation of CO by THz laser fields at zero temperature. Local control is used in Sec. \ref{sec4} to manipulate efficiently
the photodissociation of HeH$^+$. Conclusion and prospective views are given in Sec. \ref{sec5}.

\section{Formulation of optimal and local control theory with zero-area constraint}\label{sec2}
\subsection{The optimal control algorithm}\label{secopti}
We consider a quantum system whose dynamics is governed by the following Hamiltonian:
\begin{equation}
H(t)=H_0+E(t)H_1
\label{eq:Ham}
\end{equation}
where $E(t)$ is the control field. The state $|\psi\rangle$ of the system satisfies the differential equation:
\begin{equation}
i\frac{\partial |\psi\rangle }{\partial t}=H(t)|\psi\rangle ,
\label{eq:dyn}
\end{equation}
with $|\psi_0\rangle$ the initial ($t=0$) state. The units used throughout the paper are atomic units. As mentioned in the introduction, we add the physical constraint of zero-area on the control field:
\begin{equation}
A(t_f)=\int_0^{t_f}E(t)dt=0,
\label{eq:zero-area}
\end{equation}
where $t_f$ is the control (or total pulse) duration. The goal of the control problem is to maximize the following standard cost functional at time $t=t_f$:
\begin{equation}
\mathcal{J^{\rm{oc}}} = \Re [\langle \psi_f|\psi(t)\rangle] -\lambda \int_0^{t_f} E(t)^2dt,
\label{eq:Joc}
\end{equation}
where $|\psi_f\rangle $ is the target state. Other cost functionals could of course be chosen. More specifically, a monotonic optimal control algorithm can be formulated to satisfy the zero-area constraint. For such a purpose let us consider the following cost functional
\begin{equation}
  \mathcal{J^{\rm{oc}}}=\Re [\langle \psi_f|\psi(t_f)\rangle]-\mu \left[\int_0^{t_f}{E(t)dt}\right]^2-\lambda \int_0^{t_f} {[E(t)-E_{ref}(t)]^2/S(t)dt}\,,
\label{eq:newJoc}
\end{equation}
where $E_{ref}$ is a reference pulse  and $S(t)$ an envelope shape given by $S(t)
= \sin^2(\pi t/t_f)$ \cite{sunderman}. The function $S(t)$ ensures that the field is smoothly
switched on and off at the beginning and at the end of the control. The positive parameters $\mu$ and $\lambda$, expressed in a.u., weight the different parts of $\mathcal{J^{\rm{oc}}}$, which penalize the area of the field ($\mu$- term) and its energy ($\lambda$- term).
In this algorithm, we determine the field $E_{k+1}$ at step $k+1$ from the field $E_k$ at step $k$, such that the variation $\Delta \mathcal{J^{\rm{oc}}}=\mathcal{J^{\rm{oc}}}(E_{k+1})-\mathcal{J^{\rm{oc}}}(E_k)\geq 0$. At step $k$, the reference field $E_{ref}$ is taken as $E_k(t)$ and we denote by $A_k=\int_0^{t_f}E_k(t)dt$ its time-integrated area. Note that the choice $E_{ref}=E_k$ leads to a smooth transition of the control field between two iterations of the algorithm \cite{palao,bartana}.

The variation $\Delta  \mathcal{J^{\rm{oc}}}$ can be expressed as follows:
\begin{equation}
\Delta \mathcal{J^{\rm{oc}}}= \int_0^{t_f}(E_{k+1}-E_k)[\Im [\langle \chi_k |H_1|\psi_{k+1} \rangle ]-\lambda (E_{k+1}-E_k)/S(t)-2\mu A_k (E_{k+1}-E_k)]dt,
\end{equation}
$|\chi_k (t)\rangle$ is obtained from backward propagation of the target $|\psi_f \rangle$ taken as an intial state for Eq.(2) and we assume that $\int_0^{t_f}E_k(t)dt\simeq \int_0^{t_f} E_{k+1}dt$. One deduces that the choice:
\begin{equation}
E_{k+1}-E_k=\frac{S(t)\Im [\langle \chi_k |H_1|\psi_{k+1} \rangle ]}{\lambda}-(E_{k+1}-E_k)-2\frac{\mu}{\lambda}A_k S(t)
\end{equation}
ensures a monotonic increase of $\mathcal{J^{\rm{oc}}}(E_k)$. This leads to the following correction of the control field:
\begin{equation}
E_{k+1}=E_k+\frac{S(t)\Im [\langle \chi_k |H_1|\psi_{k+1} \rangle ]}{2\lambda}-\frac{\mu}{\lambda}S(t)A_k.
\label{eq:newfield}
\end{equation}
The monotonic algorithm can thus be schematized as follows:
\begin{enumerate}
\item Guess an initial control field, which at step $k$ is taken as  $E_{ref}(t) = E_k(t)$.
\item Starting from $|\psi_f\rangle$, propagate $|\chi_k (t)\rangle$ backward with $E_k(t)$
\item Evaluate the correction of the control field $E_{k+1}-E_k $ as obtained from Eq.~(\ref{eq:newfield}),
  while propagating forward the state $|\psi_{k+1}\rangle$ of the system from $|\psi_0\rangle$ with $E_{k+1}$.
\item With the new control, go to step 2 by incrementing the index $k$ by 1.
\end{enumerate}

\subsection{Local control}
The same formalism can be used to extend the local control approach to the zero-area constraint. We refer the reader to \cite{revuelocal} for a complete introduction of this method.
Using the same notations as in the previous section, we consider as Lyapunov function $\mathcal{J^{\rm{lc}}}$ the cost functional:
\begin{equation}
\mathcal{J^{\rm{lc}}}(t)=\langle \psi(t)|O|\psi(t)\rangle -\mu A(t)^2,
\label{eq:Jlc}
\end{equation}
where $O$ is any operator such that $i\frac{\partial}{\partial t}O=[H_0,O]$. The cost $\mathcal{J^{\rm{lc}}}$ is maximum at a given time $t_f$ if $\langle \psi(t_f)|O|\psi(t_f)\rangle$ is maximum and $A(t_f)=0$. The computation of the time derivative of $\mathcal{J^{\rm{lc}}}$ gives:
\begin{equation}
\dot{\mathcal{J^{\rm{lc}}}}(t)= E(t)\langle \psi(t)|\frac{[O,H_1]}{i}|\psi(t) \rangle -2\mu A(t) E(t).
\end{equation}
One can choose:
\begin{equation}
E(t)=\varepsilon  \big(\langle \psi(t)|\frac{[O,H_1]}{i}|\psi(t) \rangle -2\mu A(t)\big) ,
\label{eq:fieldlc}
\end{equation}
to ensure the monotonic increase of $\mathcal{J^{\rm{lc}}}$ from the initial state of the system $|\psi_0\rangle$. The parameter $\varepsilon$ is a small positive parameter which is adjusted to limit the amplitude of the control field.
In order to illustrate the
efficiency of the algorithm with the zero-area constraint, we introduce two
measures:
\begin{equation}
  A_\mathrm{norm} = \int_0^{t_f}{ E(t) dt} / \int_0^{t_f}{ |E(t)| dt}\ \hspace{1cm} B_\mathrm{norm} = \int_0^{t_f}{ E_\mathrm{wc}(t) dt} / \int_0^{t_f}{ E_\mathrm{woc}(t) dt}
\label{eq:Anorm}
\end{equation}
where $E_\mathrm{wc}$ and $E_\mathrm{woc}$  are the optimized field with and without
the zero-area constraint, respectively.


\section{Molecular orientation dynamics with zero-area fields}\label{sec3}
\subsection{Model system}
In this section, we present the theoretical model for the control of molecular orientation by means of THz laser pulses \cite{revuerotation1,revuerotation2,cong,fleischer,henriksen}. We consider the linear CO molecule described in a rigid rotor approximation and driven by a linearly polarized electric field $\vec{E}(t)$ along the $z$- axis of the laboratory frame. The molecule is assumed to be in its ground vibronic state. The Hamiltonian of the system is given by:
\begin{equation}
  \label{eq:hamCO}
H(t)=BJ^2-d\cos\theta E(t).
\end{equation}
The first term is the field-free rotational Hamiltonian, where $B$ is the rotational constant of the molecule and $J^2$ the angular momentum operator. The second term represents the interaction of the system with the laser field, the parameter $d$ being the permanent molecular dipole moment. The spatial position of the diatomic molecule is given in the laboratory frame by the spherical coordinates $(\theta,\phi)$. Due to the symmetry of revolution of the problem with respect to the $z$- axis, the Hamiltonian $H(t)$  depends only on $\theta$, the angle between the molecular axis and the polarization vector. Numerical values of the molecular parameters are taken as $B=1.9312~\textrm{cm}^{-1}=8.79919\times 10^{-6}~\textrm{a.u.}$ and $d=0.044~\textrm{a.u.}$. In the case of zero rotational temperature ($T=0~\textrm{K}$) which is assumed hereafter, the system is described at time $t$ by
the wave function $|\psi(t)\rangle$ and its dynamics by the time-dependent Schr\"odinger equation (Eq.~(\ref{eq:dyn})).
The dynamics is solved numerically using a third-order split operator algorithm \cite{split}.
We finally recall that the degree of orientation of the molecular system is given by the expectation value $\langle \cos\theta \rangle$.

\subsection{Orientation dynamics}
\begin{figure}[tb]
  \centering
  \includegraphics[width=0.74\linewidth]{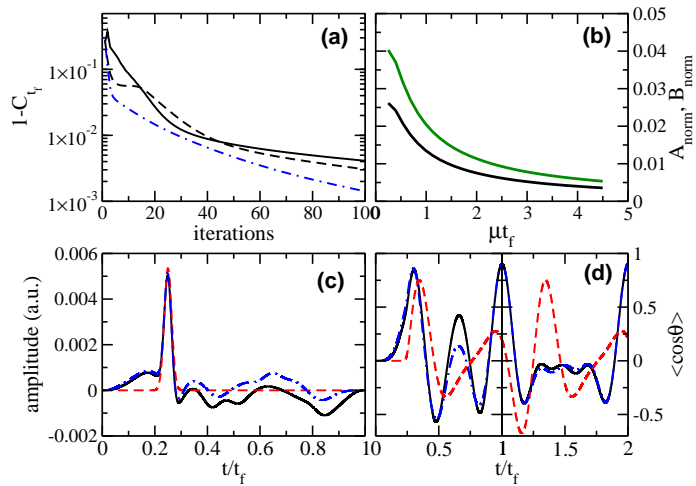}
  \caption{(Color online)
    (a) Deviation of the objective $C_{t_f}$ from unity as a function of the number of iterations for
    $\mu=0$ corresponding to the optimization without zero-area
    constraint (dashed dotted blue) and for $\mu \neq 0$ (solid black
     $\mu =  1.8/t_f$, dashed black  $\mu =  0.25/t_f$).
    (b)  Plot of $A_\mathrm{norm}$  (solid black) and $B_\mathrm{norm}$ (solid green) as a function
    of the parameter $\mu$ times the final time $t_f$.
    (c) The guess field (dashed red) and
        the optimized fields without (dotted dashed blue) and with (solid
        black) zero-area constraint.
     (d) Comparison of the time evolution of the expectation value of $\cos\theta$ with the guess field
      (dashed red) and with the fields  without (dotted dashed blue)
      and with (solid black) zero-area constraint.
    }
  \label{fig:JT_area}
\end{figure}

In this paragraph, we test the new formulation of the optimal control algorithm presented in Sec. \ref{secopti} on the specific example of CO orientation dynamics. The initial state is the ground state denoted $|0,0\rangle$ in the spherical
harmonics basis set $\{|j,m\rangle\}$. Due to symmetry, the projection $m$ of the angular momentum on the $z-$axis is a good quantum number. This property leads to the fact that only the states $|j,0\rangle$, $j\geq 0$, will be populated during the dynamics. In the numerical computations, we consider a finite Hilbert space of size $j_{max}=15$. Note that this size is sufficient for an accurate description of the dynamics considering the intensity range of the laser field used in this work. To define the cost functional, rather than maximizing the expectation value $|\langle \cos\theta\rangle |$, we focus on a target state $|\psi_f\rangle$ maximizing this value in a finite-dimensional Hilbert space $\mathcal{H}_{j_{opt}}$ spanned by the states $\{|j,0\rangle\}$ with $0\leq j\leq j_{opt}$. The details of the construction of $|\psi_f\rangle$ can be found in Refs. \cite{sugnyor1,sugnyor2}. Here, the parameter $j_{opt}$ is taken to be $j_{opt}=4$. The guess field is a Gaussian pulse of 288 fs full width at half maximum, centered at $t_0=T_{rot}/4$.
The optimization time, $t_f$, is taken to be the rotational period $T_{rot}$ of the molecule. The parameter $\lambda$ defined in the cost functional is fixed to 100. Both, with and without zero-area constraint, the optimized field achieves a fidelity higher than 99$\%$.
The results are shown in Fig.\ref{fig:JT_area}.  Note that, since the ratio $\lambda/\mu$ has the dimension of time, we found convenient to plot the evolution of the measures with respect to $\mu t_f$ (panel (b)).
The upper right panel, Fig.\ref{fig:JT_area}(a), shows the deviation from unity of the fidelity (or final objective)
\begin{equation}
C_{t_f}=|\langle \psi(t_f)|\psi_f \rangle |^2
\label{eq:objective}
\end{equation}
as a function of the number of iterations.
As could be expected, optimization with zero-area constraint
requires a large number of iterations to reach a high fidelity. For $\mu=0$, i.e. without zero-area constraint,
the final cost reaches a value of 0.99 after 30 iterations, while in the
    case for which $\mu=0.25/t_f $, this value is reached after 40 iterations.
The variations of the $A_\mathrm{norm}$ and $B_\mathrm{norm}$ with respect to
the parameter $\mu t_f$ are shown in  Fig.\ref{fig:JT_area}(b). For $\mu \ge 1/t_f$, the area of the optimized
field obtained  with the zero-area constraint is two orders of magnitude smaller
than the one obtained without the zero-area constraint. Note that
the values of $A_\mathrm{norm}$ and $B_\mathrm{norm}$ displayed in Fig.\ref{fig:JT_area}(b)
are computed with optimized fields which achieve a fidelity higher than
99$\%$. Figure \ref{fig:JT_area}(c) compares the guess (reference) field and the optimized
fields with and without the zero-area constraint. The optimized field with the
zero-area constraint shown  in Fig.~\ref{fig:JT_area}(c) is obtained with $\mu
= 4.5/t_f$. The optimized fields are similar to the guess field at the beginning of the control
and become slightly different for $t/t_f>0.2$. Figure \ref{fig:JT_area}(d) corresponds to the time evolution of
$\langle\cos\theta\rangle$ during two rotational periods (one with the field and one without).
It compares the dynamics of $\langle\cos\theta\rangle(t)$
induced by the three fields shown in the lower left panel. Clearly, the dynamics with the guess field is very different from the one obtained with the optimized fields, which have similar features.

\section{Photodissociation}\label{sec4}
\subsection{Model system}
We consider the photodissociation of HeH$^+$ through the singlet $^1\Sigma$ excited states coupled by numerous nonadiabatic interactions.
A local control strategy has recently been applied to find a field able to enhance the yield in two specific dissociation
channels either leading  to He$^*(2s)$, or to He$^*(2p)$  \cite{LocalHeH}. The diatomic system is described by its reduced mass $m$ and the internuclear distance $R$.
Due to the short duration of the pulses (as compared with the rotational period) a frozen rotation approximated is valid. In addition, the molecule is assumed to be pre-aligned along the $z$-direction of the laboratory frame. The initial state ${|\psi_0 \rangle}$ is the vibrationless ground electronic adiabatic state. Dynamics is performed in the diabatic representation. The total Hamiltonian is written as:
\begin{equation}
H = H_0  - \sum_{i,j=1}^N M^d_{ij}(R) E(t)  =
\sum_{i,j=1}^N |\varphi^d_{i}\rangle  (T_{ij}\delta_{ij}+V^d_{ij}(R)) \langle \varphi^d_{j} |- \sum_{i,j=1}^N M^d_{ij}(R) E(t)
\label{eq:hamHeH+}
\end{equation}
where $H_0$ is the field free Hamiltonian, $N$ is the number of diabatic electronic states $| \varphi^d_{i} \rangle$ under consideration
and $T_{ii} = - (1/2m )\partial^2 /\partial R^2  $. All the adiabatic potential energy curves and the nonadiabatic couplings
$F_{ij}=\langle \varphi^a_i|\partial/\partial R|\varphi^a_j\rangle$ have been computed in Ref.~\cite{HeH}.  $M^d_{ij}(R)$ are the
diabatized dipole transition matrix elements. We shall consider only parallel transitions among the singlet  $^1\Sigma$ states
induced by the $d_z$ dipole operator by assuming the internuclear axis pointing along the $z$-direction.
The adiabatic-to-diabatic transformation matrix $D$ has been derived by integrating $\partial D/ \partial R +FD = 0$ from the
asymptotic region where both representations coincide.

\subsection{Dissociation control}
The local control of a nonadiabatic dissociation requires a careful choice of the operator $O$ referred to in Eq.~(\ref{eq:Jlc})
since it has to commute with the field free Hamiltonian \cite{LocalHeH}. In the nonadiabatic case, the projectors on
either adiabatic or diabatic states are thus not relevant as they do not commute with this Hamiltonian due to kinetic or
potential couplings respectively. This crucial problem can be overcome by using projectors on eigenstates of $H_0$, i.e.
on scattering states correlating with the controlled exit channels. In this example, the  operator $O$ takes the form:
\begin{equation}
O  =  \sum_{p \in S} \int dk |\xi_p^-(k)\rangle \langle \xi_p^-(k) | \label{newproj}
\end{equation}
where $S$ represents the two channels leading to the target He$^*(2s,2p)$ fragments, the objective being $\langle \psi(t)|O|\psi(t)\rangle$. The local control field now reads
\begin{equation}
E(t)  =   \eta \Im( \sum_{p \in S} \int dk \langle \psi(t) | \xi_p^-(k)\rangle \langle \xi_p^-(k) |d_z | \psi(t) \rangle ) -2 \mu A(t)
\end{equation}
involving two adjustable parameters $\eta$ and $\mu$.
\begin{figure}[tb]
 \centering
 \includegraphics[width=0.74\linewidth]{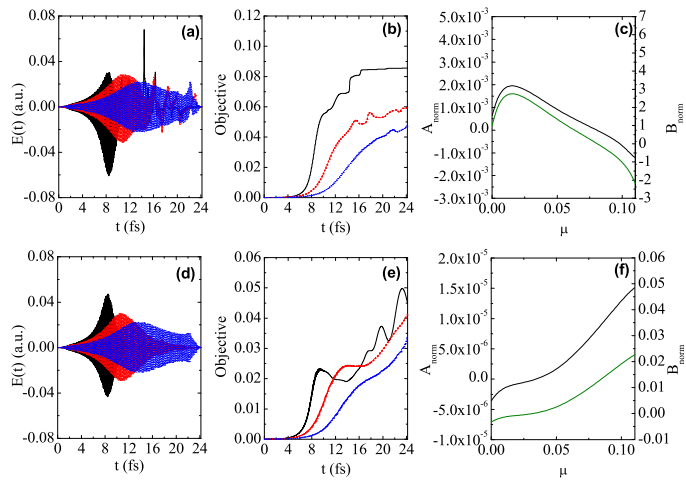}
 \caption{(Color online)
    (a) Fields obtained by the local control of the photodissociation, solid black:
  without constraint $\mu =0$, dashed red: $\mu$ = 0.05, dotted blue: $\mu$ = 0.10. The parameter $\mu$ is expressed in atomic units. (b) Objectives of
  the local control measured by the total He$^*(2s,2p)$ yield (same color code than in (a)).
(c) $A_{norm}$ (solid black) and $B_{norm}$ (solid green) as a function of $\mu$.
(d) Fields after filtration of the low frequencies of the spectrum removing the static Stark component.
(e) Objectives with the filtered fields. (f) $A_{norm}$ and $B_{norm}$ (same color code than in (c)) as a function
of $\mu$ for the filtered fields. }
      \label{fig:HeH_area}
\end{figure}

The ingoing scattering states $|\xi_p^-(k)\rangle$ are estimated using a time-dependent approach based on M{\o}ller-operators
\cite{Tannor} as derived in Ref. \cite{LocalHeH}.
This control strategy remains local in time but can preemptively account for nonadiabatic transitions that occur later
in the dynamics. The photodissociation cross section \cite{Sodoga} shows that there is no spectral range where
the He$^*(2s,2p)$ dissociation channels dominate. The local control finds a very complicated electric field which
begins by an oscillatory pattern followed after 10 fs by a strong complex positive shape whose area is obviously not
zero (see Fig.~\ref{fig:HeH_area}(a)). We therefore choose this example to check the efficiency of the zero-area constraint algorithm.
We use $ \eta =$ 4.2 and different values of $\mu$ to assess the effect of the zero-area constraint on the control field.
Figure \ref{fig:HeH_area}(a) shows the dramatic evolution of the pulse profile for two values of $\mu$ (0.05 and 0.10).
The main change is the suppression of the erratic positive structure which can be interpreted as a Stark field.
However, as shown in Fig.~\ref{fig:HeH_area}(c) the variation of $B_{norm}$ is not monotonic and the best correction is actually obtained for a given value of $\mu$ around $\mu = 0.08$, leading to almost zero value for $B_{norm}$. Close to the optimal $\mu$, one observes a clear improvement of the zero-area
constraint, but at the price of a decrease of the objective (from $8\%$ without constraint to $4\%$ with $\mu = 0.10$) as
shown in Fig.~\ref{fig:HeH_area}(b). However, a brute force strategy by filtration of near-zero frequencies already
provides a large correction to the non vanishing area while also decreasing the yield.
The field with $\mu =0$ and after filtration of the low frequencies is shown in black line in Fig.~\ref{fig:HeH_area}(d) and
the corresponding yield is given in Fig.~\ref{fig:HeH_area}(e). The objective is divided by a factor 2.
We also filter the fields generated by the constrained control with $\mu$ equal to 0.05 and 0.10 which improves even further the zero-area
constraint. The measures $B_{norm}$ evaluated with the filtered fields (Fig.~\ref{fig:HeH_area}(f)) are smaller than the unfiltered
ones by about two orders of magnitude, while the objectives are reduced by about a factor 1.5, as seen by comparing Fig.~\ref{fig:HeH_area}(e)
and Fig.~\ref{fig:HeH_area}(b).

\section{Conclusion}\label{sec5}
We have presented new formulations of optimization algorithms in order to design control fields with zero area.
The procedure is built from the introduction of a Lagrange multiplier aiming at penalizing the area of the field. The value of the corresponding parameter is chosen to express the relative weight between the area and the other terms of the cost functional, i.e. the projection onto the target state and the energy of the electric field.
The efficiency of the algorithms is exemplified on two key control problems of molecular dynamics: namely, the molecular orientation
of CO and the photodissociation of HeH$^+$.

From a theoretical view point, the zero-area constraint is crucial since all electromagnetic fields which are
physically realistic must fulfill this requirement. Surprisingly, this physical requirement has been considered, up to date,
by only very few works in quantum control. As such, this work can be viewed as a major advance of the state-of-the-art of
optimization algorithms that could open the way to promizing future prospects.
\section{Acknowledgments}
S. V. acknowledges financial support from the Fonds de la Recherche Scientifique (FNRS) of Belgium. Financial
support from the Conseil R\'egional de Bourgogne and the
QUAINT coordination action (EC FET-Open) is gratefully
acknowledged by D. S. and M. N..


\begin{thebibliography}{23}
\markboth{Taylor \& Francis and I.T. Consultant}{Journal of Modern Optics}

\bibitem[1]{pont}
Pontryagin, L. et al. {\em Mathematical theory of optimal processes}; Mir, Moscou, 1974.

\bibitem[2]{bryson} Bryson, A. E.; Ho, Y.-C. {\em Applied optimal control}; Taylor \& Francis, New York London, 1975.

\bibitem[3]{rice}
Rice, S.; Zhao, M. {\em Optimal control of molecular dynamics}; Wiley, New York, 2003.

\bibitem[4]{brumer}
Shapiro, M.; Brumer, P. {\em Principles of quantum control of molecular processes}; Wiley, New York, 2003.

\bibitem[5]{revuerabitz}
Brif, R.C.C.; Rabitz, H. {\em New. J. Phys.} {\bf 2010}, {\em 12} 075008.

\bibitem[6]{glaser} Skinner, T. E.; Reiss, T. O.; Luy, B.; Khaneja, N.; Glaser, S. J. {\em J. Magn. Reson.} {\bf 2003}, {\em 163} 8.

\bibitem[7]{krotov} Soml\'oi, J.; Kazakovski, V. A.; Tannor, D. J. {\em Chem. Phys.} {\bf 1993}, {\em 172} 85.

\bibitem[8]{reich}
Reich, D. M.; Ndong, M.; Koch, C. P. {\em J. Chem. Phys.} {\bf 2012}, {\em 136} 104103.

\bibitem[9]{zhu} Zhu, W.; Botina, J.; Rabitz, H. {\em J. Chem. Phys.} {\bf 1998}, {\em 108} 1953.

\bibitem[10]{maday} Maday, Y.; Turinici, G. {\em J. Chem. Phys.} {\bf 2003}, {\em 118} 8191.

\bibitem[11]{revuelocal}
Engel, V.; Meier, C.; Tannor, D. {\em Adv. Chem. Phys.} {\bf 2009}, {\em 141} 29.

\bibitem[12]{revuealessandro} D'Alessandro, D. {\em Introduction to quantum control and dynamics}; Boca Raton, FL, CRC Press 2007.

\bibitem[13]{sugawara} Sugawara, M. {\em J. Chem. Phys.} {\bf 2003}, {\em 118} 6784.

\bibitem[14]{wang} Wang, X.; Schirmer, S. G.; {\em Phys. Rev. A} {\bf 2009}, {\em 80} 042305.

\bibitem[15]{spectrum1}
Gollub, C.; Kowalewski, M.; Vivie-Riedle R. {\em Phys. Rev. Lett.} {\bf 2008}, {\em 101} 073002.

\bibitem[16]{spectrum2}
Lapert, M.; Tehini, R.; Turinici, G.; Sugny, D. {\em Phys. Rev. A} {\bf 2009}, {\em 79} 063411.

\bibitem[17]{nonlinear1}
Ohtsuki, Y.; Nakagami, K. {\em Phys. Rev. A} {\bf 2008}, {\em 77} 033414.

\bibitem[18]{nonlinear2}
Lapert, M.; Tehini, R.; Turinici, G.; Sugny, D. {\em Phys. Rev. A} {\bf 2008}, {\em 78} 023408.

\bibitem[19]{kobzar}
Kobzar, K.; Skinner, T. E.; Khaneja, N.; Glaser, S. J.; Luy, B. {\em J. Magn. Reson.} {\bf 2004}, {\em 170} 236.

\bibitem[20]{zhang} Zhang, Y.; Lapert, M.; Braun, M.; Sugny, D.; Glaser, S. J. {\em J. Chem. Phys.} {\bf 2011}, {\em 134} 054103.

\bibitem[21]{you}
You, D.; Bucksbaum, P. H. {\em J. Opt. Soc. Am. B} {\bf 1997}, {\em 14} 1651.

\bibitem[22]{rabitz}
Liao, S.-L.; Ho, T.-S.; Rabitz, H.; Chu, S.-I. {\em Phys. Rev. A} {\bf 2013}, {\em 87} 013429.

\bibitem[23]{ortigoso}
Ortigoso, J. {\em J. Chem. Phys.} {\bf 2012}, {\em 137} 044303.

\bibitem[24]{sugny}
Sugny, D.; Keller, A.; Atabek, O.; Daems, D.; Gu\'erin, S.; Jauslin, H. R. {\em Phys. Rev. A} {\bf 2004}, {\em 69} 043407.

\bibitem[25]{lapert}
Lapert, M.; Sugny, D. {\em Phys. Rev. A} {\bf 2012}, {\em 85} 063418.

\bibitem[26]{sunderman}
Sunderman, K.; de Vivie-Riedle, R. {\em J. Chem. Phys.} {\bf 1999}, {\em 110} 1896.

\bibitem[27]{palao} Palao, J. P.; Kosloff, R. {\em Phys. Rev. Lett.} {\bf 2002}, {\em 89} 188301.

\bibitem[28]{bartana} Bartana, A.; Kosloff, R.; Tannor, D. J. {\em Chem. Phys.} {\bf 2001}, {\em 267} 195.

\bibitem[29]{revuerotation1}
Seideman, T.; Hamilton, E. {\em Adv. At. Mol. Opt. Phys.} {\bf 2006}, {\em 52} 289.

\bibitem[30]{revuerotation2}
Stapelfeldt, H.; Seideman, T. {\em Rev. Mod. Phys.} {\bf 2003}, {\em 75} 543.

\bibitem[31]{cong}
Shu, C. C.; Yuan, K. J.; Hu, W. H.; Cong S. L. {\em J. Chem. Phys.} {\bf 2010}, {\em 132}, 244311.

\bibitem[32]{fleischer}
Fleischer, S.; Zhou, Y.; Field, R. W.; Nelson, K. A. {\em Phys. Rev. Lett.} {\bf 2011}, {\em 107} 163603.

\bibitem[33]{henriksen}
Shu, C.-C.; Henriksen, N. E. {\em Phys. Rev. A} {\bf 2013}, {\em 87} 013408.

\bibitem[34]{viellard}
Viellard, T.; Chaussard, F.; Sugny, D.; Lavorel, B.; Faucher, O. {\em J. Raman Spec.} {\bf 2008}, {\em 39} 694.

\bibitem[35]{split}
Feit, M. D.; Fleck, J. A.; Steiger, A. {\em J. Comput. Phys.} {\bf 1982}, {\em 47} 412.

\bibitem[36]{sugnyor1}
Sugny, D.; Keller, A.; Atabek, O.; Daems, D.; Dion, C. M.; Gu\'erin, S.; Jauslin, H. R. {\em Phys. Rev. A} {\bf 2005}, {\em 71} 063402.

\bibitem[37]{sugnyor2}
Sugny, D.; Keller, A.; Atabek, O.; Daems, D.; Dion, C. M.; Gu\'erin, S.; Jauslin, H. R. {\em Phys. Rev. A} {\bf 2005}, {\em 72} 032704.

\bibitem[38]{LocalHeH}
Bomble, L.; Chenel, A.; Meier, C.; Desouter-Lecomte M. {\em J. Chem. Phys.} {\bf 2011}, {\em 134}, 204112.

\bibitem[39]{HeH}
Loreau, J.; Palmeri, P.; Quinet, P.; Li\'evin, J.; Vaeck, N.  {\em J. Phys. B: At. Mol. Opt. Phys.} {\bf 2010}, {\em 43} 065101.

\bibitem[40]{Tannor}
Tannor, D. J.;  {\em Quantum Mechanics: A Time Dependent Perspective} {\bf 2007}, University Science Book: Sausalito, CA.

\bibitem[41]{Sodoga}
Sodoga, A.; Loreau, J.; Lauvergnat, D.; Justum, Y.; Vaeck, N.; Desouter-Lecomte, M.  {\em Phys. Rev. A} {\bf 2009}, {\em 80} 033417.
\end{thebibliography}
\end{document}